# Delivery of focused short pulses through a multimode fiber

**Edgar Emilio Morales Delgado[1]\*, Salma Farahi[1,2], Ioannis Papadopoulos[2], Demetri Psaltis[2] and Christophe Moser[1]**

[1]*Laboratory of Applied Photonics Devices, School of Engineering, École Polytechnique Fédéral de Lausanne (EPFL), Station 17, 1015, Lausanne, Switzerland*

[2]*Laboratory of Optics, School of Engineering, École Polytechnique Fédéral de Lausanne (EPFL), Station 17, 1015, Lausanne, Switzerland*

*\*edgar.moralesdelgado@epfl.ch*

**Abstract:** Light propagation through multimode fibers suffers from spatial distortions that lead to a scrambled intensity profile. In previous work, the correction of such distortions using various wavefront control methods has been demonstrated in the continuous wave case. However, in the ultra-fast pulse regime, modal dispersion temporally broadens a pulse after propagation. Here, we present a method that compensates for spatial distortions and mitigates temporal broadening due to modal dispersion by a selective phase conjugation process in which only modes of similar group velocities are excited. The selectively excited modes are forced to follow certain paths through the multimode fiber and interfere constructively at the distal tip to form a focused spot with minimal temporal broadening. We demonstrate the delivery of focused 500 fs pulses through a 30 cm long step-index multimode fiber. The achieved pulse duration corresponds to approximately 1/30th of the duration obtained if modal dispersion was not controlled. Moreover, we measured a detailed two-dimensional map of the pulse duration at the output of the fiber and confirmed that the focused spot produces a two-photon absorption effect. This work opens new possibilities for ultra-thin multiphoton imaging through multimode fibers.

**1. Introduction**

Multimode optical fibers have garnered considerable attention lately, mainly because of the large information capacity that they provide, linked to the number of supported propagation modes. In this context, image transmission and image formation through multimode fibers have transformed them into a new paradigm of miniature imaging devices [1-4] as well as potential high throughput channels in optical communications [5,6].

The transmission of ultrashort pulses through a multimode fiber suffers from mode mixing which leads to a scrambling of the spatial profile and from modal dispersion that causes a broadening of the temporal profile of the propagating pulse. Although the pattern emerging from the multimode fiber looks like a random field both in space and time, the process is in fact linear and deterministic. Hence, polychromatic light propagation through a multimode fiber is a reversible phenomenon.

Modern optical communications mainly rely on single mode fibers, which provide waveguiding with a well-defined spatial profile. However, pulse duration is affected by chromatic dispersion and optical non-linear effects. Several methods have been developed to counteract temporal spreading to recover a pulse of similar duration as the original. For example, temporal pulse shaping methods and phase conjugation have been used successfully [7-9].

When a multimode fiber is used, additional temporal broadening takes place due to modal dispersion, i.e., the different modes propagating in the fiber have different group velocities. Contrary to the case of single mode fibers for which the spatial profile is maintained after propagation, the pattern emerging from a multimode fiber does not resemble the pattern launched at the input as explained before. Adaptive methods have been proposed to compensate for modal dispersion in telecommunications in order to deliver femtosecond pulses through multimode fibers but no simultaneous spatial control has been achieved [10,11].

Regarding the spatial domain, in the monochromatic case, optical phase conjugation was first suggested [12,13,14] and demonstrated as a means to undo the modal scrambling and transmit images through multimode fibers. More recently, the field was revisited in the digital domain, where several methods such as iterative algorithms [15,16], transmission matrix measurement [2,3,17] or digital phase conjugation [18] have been successfully used to spatially focus light and image through multimode optical fibers [1-4]. However, the control of both the spatial profile and the temporal duration of pulses through multimode fibers have not yet been demonstrated.

Interestingly, spatial wavefront shaping techniques have been proposed for focusing through a scattering medium in space and time [19,20]. These methods rely on iterative algorithms that optimize a signal, such as a two-photon signal, that strongly depends on the simultaneous spatial and temporal focusing of a light pulse. They are based on the fact that light propagating through such scattering media is re-radiated at every scattering event, creating different light paths. Each path is associated with a respective time delay. This spatio-temporal relationship allows the correction of both temporal and spatial distortions using only spatial degrees of wavefront control.

For information transmission, delivery of an arbitrary spatial distribution of ultrashort light pulses could enable an efficient spatial division multiplexing of orthogonal communication channels with limited dispersion [6]. In the field of imaging, the ability to deliver and digitally scan focused ultrashort pulses would allow multiphoton lensless imaging, which has only been accomplished using fiber bundles [21].

In the present paper we demonstrate for the first time the delivery of spatially focused 500 fs optical pulses through a 200 μm core diameter, 30 cm long step-index multimode fiber. Our approach minimizes modal dispersion in the multimode fiber by selectively counter-propagating only a group of modes of similar group velocities, hence limiting dispersion. Specifically, as an initial step described in detail in section 2, we couple light on the distal side of an optical fiber and use time-gated interferometry [22] and digital holography to characterize the optical field at the proximal side. Then, we use digital phase conjugation (DPC) to reconstruct, on the proximal side, a field that corresponds to a time-gated set of modes. The reconstructed field counter-propagates through the fiber generating an ultrashort focused pulse at the distal end of the fiber. Since the power of the counter-propagated field is spatially spread on the various excited modes, our approach does not introduce nonlinear effects on the delivered focused pulse. In section 3, we explain a method to map across the facet of the multimode fiber core the pulse duration of the focused spot and we demonstrate that a two-photon effect can be produced at the location of the spot. In section 4, we show that the location of the focused spot can be changed by digital means. Additionally, we show how to send multiple focused spots separated in time.

**2. Selective mode sampling and reconstruction**

The beam from a pulsed laser source (ORIGAMI-15 from OneFive; λ=1550 nm; spectral width $σ_λ$=15.4 nm; pulse energy = 2 nJ) is amplified by a custom built Chirped Pulse Amplification unit (CPA), whose output has a temporal duration of 440 fs, with a repetition rate of 40 MHz and a maximum average power of 1.2 W, yielding a maximum pulse energy of 20 nJ. The selective DPC technique consists of two separate steps: a calibration step for mode selection and a reconstruction step.

*2.1 Calibration – mode selection*

Initially the output of the CPA unit is focused, at the distal side, close to the facet of the multimode fiber (core diameter = 200 μm, NA = 0.39, length = 0.3 m, number of supported modes M≈$10^3$) as illustrated in Fig. 1. The M excited modes propagate with different propagation constants and acquire different phase shifts, leading to a scrambled amplitude and phase field at the proximal side of the fiber. This field can be expressed as a linear combination of the supported modes of the fiber (one polarization) with different scaling coefficients and phase factors around a carrier frequency:

$$E_{out}(x,y,t) = \sum_{m=1}^{M} a_m(t) \psi_m(x,y) e^{j\phi_m(x,y,t)} e^{-j\omega_0 t}, \qquad [1]$$

where $a_m(t)$ is the time dependent scaling coefficient of mode m, $\psi_m(x,y)$ is the amplitude of the wave function of mode m, $\phi_m(x,y,t)$ is the corresponding modal phase, and $\omega_o$ is the carrier frequency. In this form, propagation loss, material dispersion and the phase shift introduced during propagation are included in the terms $a_m(t)$ and $\phi_m(x,y,t)$. Figure 2 (a) shows the intensity of the field $E_{out}(x,y,t)$ as measured by Camera 1.

To characterize the field $E_{out}(x,y,t)$ we interfere it off-axis with a reference pulse at the plane of Camera 1 as shown in Fig. 1. The reference pulse can be described as,

$$E_{ref}(x,y,t-\tau) = a_{ref}(x,y,t-\tau) e^{-j\omega_0(t-\tau)} \qquad [2]$$

where $a_{ref}(x,y,t-\tau)$ is the complex amplitude envelope of the reference. Because the duration of the reference pulse is much shorter that the duration of the pulse at the output of the fiber, $E_{out}(x,y,t)$, we can write that $a_{ref}(x,y,t) \approx a_{ref}(x,y) \cdot \delta(t)$. The reference beam acts as a sampling window in time. The reference beam can be translated in time, expressed by the factor $\tau$ in Eq. [2], using the mirror M5 of Fig. 1, which is mounted on a translation stage with a temporal resolution of 0.33 fs. For each delay $\tau$, the reference beam samples a different time slice of the output field by capturing a digital hologram, which is recorded on the Camera 1, in an off-axis configuration. For the remainder of the paper, a time-sampled part of the output field at a given delay time will be identified with the time variable $\tau_1$, while all other time dependencies will be represented by t.

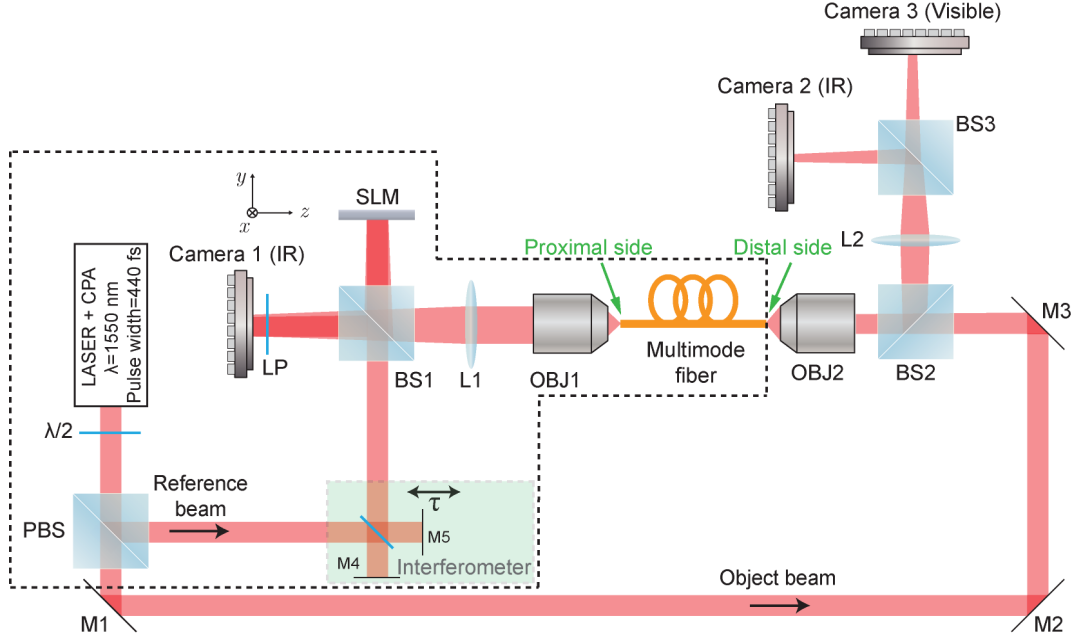

Fig. 1. Experimental Setup. *Calibration step.* The beam from the CPA unit is divided by a polarizing beam splitter PBS into a reference and an object beam. The object beam is coupled into the multimode fiber by a 20X microscope objective OBJ2. The output of the fiber is imaged on the infrared Camera 1 by a 20X microscope objective (OBJ1) and the lens (L1), f=150 mm, where it is interfered with the reference beam obtained by reflection from the beam splitter BS1. For each delay $\tau$, a digital hologram is recorded. *Reconstruction step.* The time-sampled field is reconstructed by the reference and phase conjugated using a spatial light modulator SLM. The reconstruction is imaged on the fiber by the lens L1 and the 20X microscope objective OBJ1. The reconstructed field counter-propagates generating the short focused spot on the distal side of the fiber. This spot is imaged on Camera 2 using a 4f system (OBJ2 and lens L2, f=300 mm). Moreover, the spatio-temporal duration of the phase conjugated spot and its surrounding background is measured on each pixel of a silicon-based detector (Camera 3) using second order (interferometric) autocorrelation, by introducing on the reference the collinear time-delayed replicas required for this measurement, using the Michelson interferometer. The non-linearity in the second order autocorrelation is a two-photon process occurring in the silicon camera. The dashed polygon encloses a possible embodiment of an imaging device based on our method.

The intensity of the recorded digital hologram can be expressed as a superposition of three terms: One dc term and two terms containing the phase information of the field $E_{out}(x,y,t)$ [23]. The angle between the reference beam and the object beam is adjusted to obtain an adequate separation of the interference terms in the Fourier space. The spatial Fourier transform of the digital hologram is calculated and the interference term, containing the phase information, is spatially filtered. The field power as a function of time is measured by integrating over the whole computed filtered term for each delay $\tau_1$, which gives a measure of the time duration of the pulse at the proximal side of the fiber (Fig. 2 b). Then, the inverse Fourier transform of the filtered term is calculated for each delay $\tau_1$, leading to a digital reconstruction of the phase and amplitude of the field. Therefore, the time sampled version of the field $E_{sampled}(x,y,\tau_1)$, at the plane of Camera 1 is given by,

$$E_{sampled}(x,y,\tau_1) = \sum_{m=M_a}^{m=M_b} a_m(\tau_1)\psi_m(x,y)e^{j\phi_m(x,y,\tau_1)}e^{-j\omega_0\tau_1} \,, \qquad [3]$$

where $M_a$ and $M_b$ are the first and last sampled modes of the set of modes that arrive within the temporal duration of the reference envelope.

With a reference beam of temporal duration of 440 fs, the number of modes sampled by the reference is $M_b - M_a$ =42 (one polarization). This value is obtained by taking into account the mode velocity of the M modes that propagate through the multimode fiber [24]. Specifically, we calculated the number of modes that arrive within the temporal duration of the reference. Figure 2 (c-f) shows the time-gated amplitude of the field $E_{sampled}(x,y,\tau_1)$ at time delays $\tau_1$ =2.9 ps, $\tau_1$ =7.7 ps, $\tau_1$ =13.3 ps, and $\tau_1$ =15.1 ps respectively.

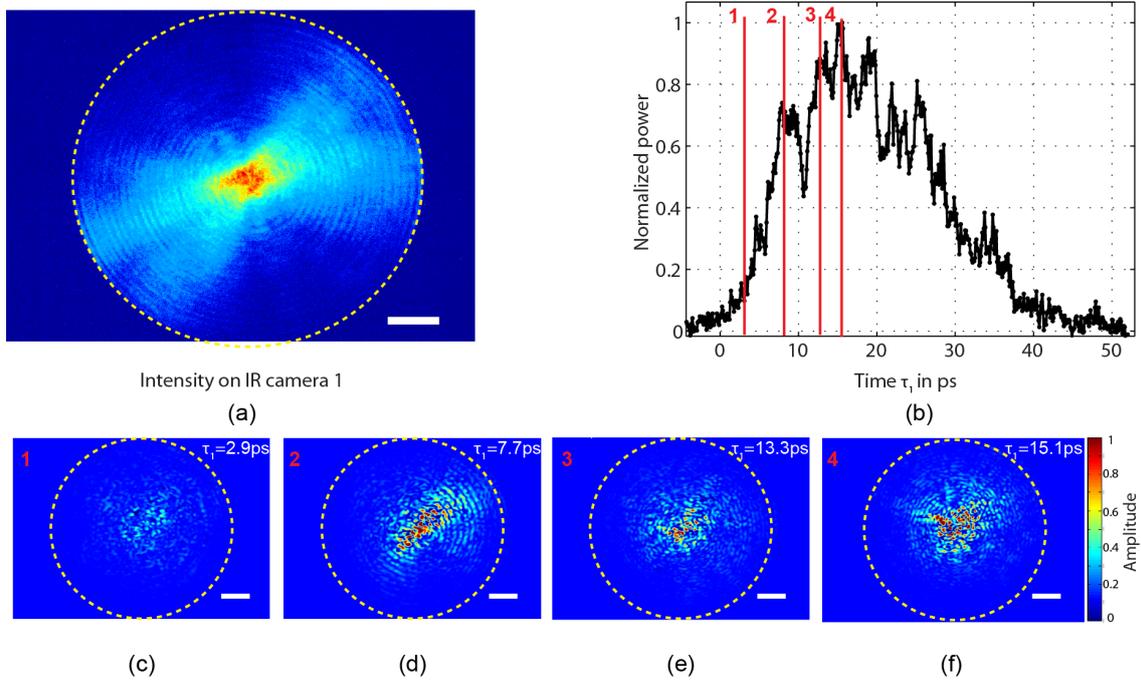

Fig. 2. Propagation and characterization of an ultrashort pulse through a multimode fiber. (a) Optical intensity as seen on the proximal end (Camera 1) containing the superposition of the excited modes arriving at all times. (b) Normalized optical power of (a) over the whole area of the Camera 1 as a function of time. (c)-(f) Time-gated snapshots of the sampled field (Eq. [3]) taken at times $\tau_1$ =2.9 ps, $\tau_1$ =7.7 ps, $\tau_1$ =13.3 ps, and $\tau_1$ =15.1 ps respectively. Scale bars are 25 μm. Dashed circles indicate the edge of the multimode fiber core.

*2.2 Modal control – delivery of short focused pulses*

By phase conjugating each one of the sampled fields shown in Fig. 2 (c-f), a focused pulse is expected to be generated at the distal side. Experimentally, the computed phase of each sampled field is conjugated and loaded serially onto a Spatial Light Modulator (Holoeye PLUTO-TELCO-013-C, phase only SLM, 1920x1080 pixels, phase range from 0 to 2π, pixel pitch 8 μm). The reference beam (Fig. 1) illuminates the SLM. The back-propagating field retraces its way through the multimode fiber to the distal end. Note that in this reconstruction step, the object beam is blocked.

The spatial amplitudes of the phase conjugated spots, measured with the infrared Camera 2, are shown in Fig. 3 (a-d), and their temporal profiles in Fig. 3 (e) for different phase conjugated fields corresponding to digital holograms recorded at time delay $\tau_1$ =2.9 ps, $\tau_1$ =7.7 ps, $\tau_1$ =13.3 ps, and $\tau_1$ =15.1 ps. The temporal profiles were obtained by second order interferometric autocorrelation averaged within the spatial FWHM of the spot (see details in section 3). The measured pulse widths are 0.50 ps, 0.80 ps, 1.11 ps and 0.54 ps respectively. The spot sizes, measured as the FWHM of the intensity line profile passing through the center of the intensity focus, are 10.5 μm, 6.3 μm, 5.7 μm and 6 μm respectively.

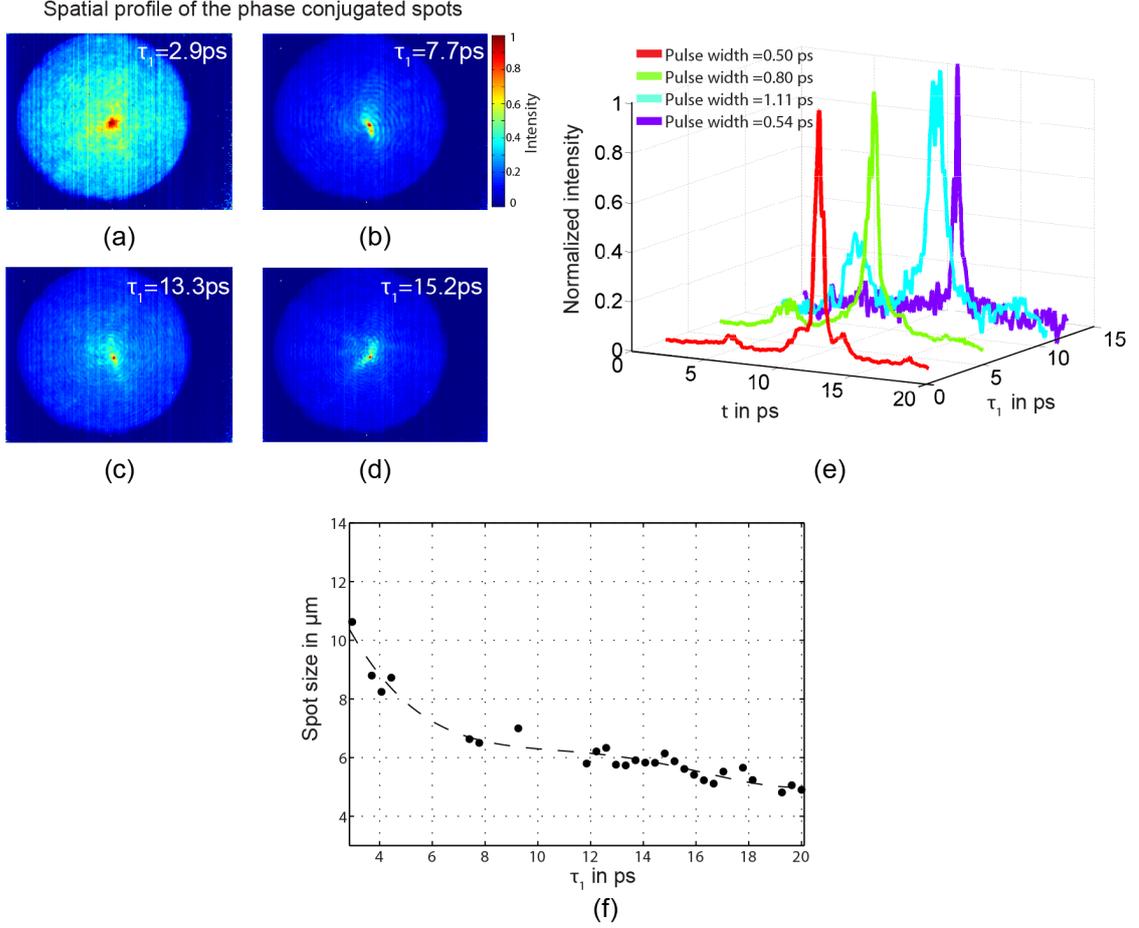

Fig. 3. Spatio-temporal characterization of the reconstructed phase conjugated spot. (a-d) Intensity of the spatial profile measured with Camera 2 and (e) temporal profile of the phase conjugated spots generated from the reconstructed holograms of Fig. 2(c-f) taken at times $\tau_1$ =2.9 ps, $\tau_1$ =7.7 ps, $\tau_1$ =13.3 ps and $\tau_1$ =15.1 ps respectively. $\tau_1$ identifies the time at which the hologram was recorded and t is the time dependence of the intensity autocorrelation trace of the phase conjugated spot. (f) Size of the phase conjugated spot as a function of $\tau_1$. Points represent experimental data and the solid curve a polynomial fit.

The evolution of the spot size of the phase conjugated spot as a function of the time $\tau_1$ at which the hologram was recorded is shown in Fig. 3 (f). During the selective DPC process, the original field is reconstructed with only 42 time-sampled modes (out of the 10'000 modes supported by the fiber). Hence, the original spot cannot be reconstructed with high fidelity. We observe that phase conjugation of higher order modes, corresponding to a larger delay $\tau_1$, produces a smaller phase conjugated spot as expected. Although the spot size decrease as a function of $\tau_1$, the temporal duration of the focus increases (see appendix A). We show that the selection of the higher order modes gives the highest spot intensity. The highest peak intensity measured in our experiment was 125 MW/cm$^2$ realized with a spot waist of 5.1 μm corresponding to the excitation of high order modes sampled at a delay $\tau_1$ = 20 ps.

## 3. Two-dimensional mapping of pulse duration

In order to obtain insights into the spatial distribution of the pulse duration around the phase conjugated spot at the distal end of the fiber, we use a silicon-based camera (Camera 3) sensitive to half the excitation wavelength to spatially resolve a second order interferometric autocorrelation. This is performed by introducing on the reference two time-delayed replicas with the Michelson interferometer shown in Fig. 1. The two paths are aligned by ensuring that the speckle pattern on Camera 2 does not change when the delay is varied. The intensity of the interferometric autocorrelation trace is measured for each camera pixel and for each position $\tau$ of the interferometer. The temporal duration of the phase conjugated field imaged on the Camera 3 is calculated at each pixel assuming a Gaussian pulse shape. To our knowledge, this is the first spatio-temporal characterization of an optical field propagated through a multimode fiber.

Fig. 4 provides a comparison of the 2D intensity and pulse duration for the case when 1. the excitation comprises a large number of fiber modes (Fig. 4 (a,c)) and 2. the excitation comprises a selected number of modes (Fig. (b,d)) using the DPC method proposed.

Fig. 4 (a) shows the intensity (measured with infrared camera 2) at the distal side of the multimode fiber generated by presenting on the SLM a set of plane waves propagating in all directions within the acceptance angle of the multimode fiber, hence exciting a large number of fiber modes.

Fig. 4 (b) shows the intensity (measured with infrared camera 2) on the distal side when a group of modes is counter-propagated using the proposed DPC method. The ratio of the peak intensity to the average background is 16. This implies that the number of phase conjugated modes is enough to generate a focused spot 16 times more intense than the background intensity. The size of the phase conjugated spot is 7 μm.

Fig. 4 (c) and (d) show the map of temporal duration corresponding to Fig. 4 (a) and (b) respectively. The average temporal duration when all fiber modes are excited is 15.2 ps (Fig. 4 (c)), which is large because modal dispersion is present. For the DPC case (Fig. 4 (d)), modal dispersion is highly suppressed, and there is a short temporal duration of 500 fs only at the location of the DPC focus. Around the focus, the temporal duration is broad (on average 10 ps) due to undesired excitation of other modes, which are subject to modal dispersion. However, these modes carry much less power (only 6% of the power of the focus). Fig. 4 (e) shows the comparison between the average temporal duration obtained when many fiber modes are excited (black curve) and when DPC is used (blue curve). Our DPC method significantly reduces modal dispersion. The broadening from the original pulse (440 fs) to the delivered pulse using DPC (500 fs) is due to material dispersion. Both pulses shown in Fig. 4 (e) contain the same energy, but the one with reduced modal dispersion has a larger intensity, which is enough to produce two-photon phenomena as shown in Fig. 5.

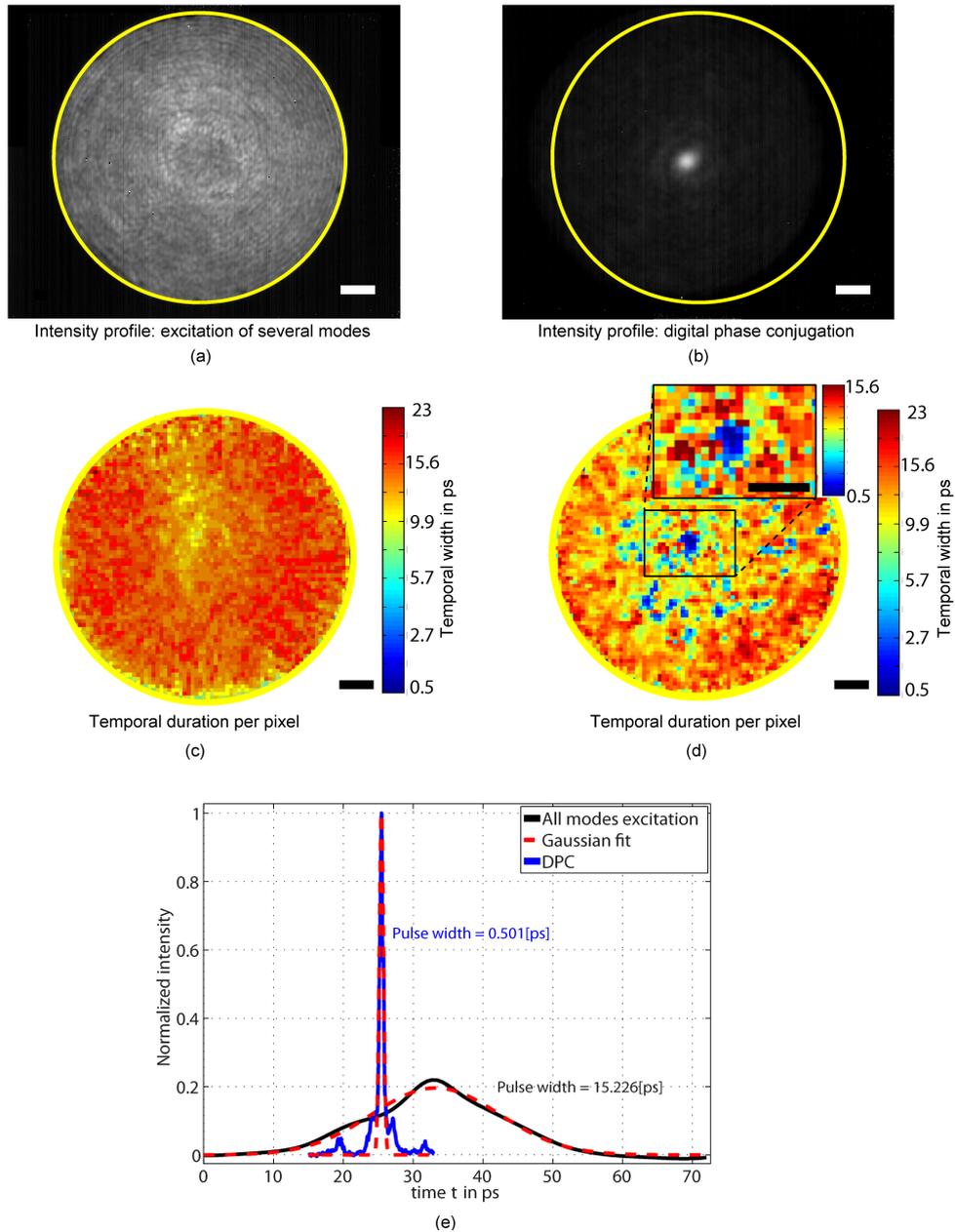

Fig. 4. Comparison between the excitation of a large number of modes and the selective DPC method. (a) Intensity when many fiber modes are excited. (b) Intensity of a phase conjugated spot generated using DPC. The size of the spot is 7 μm and is 16 times more intense that the background. (c) and (d) are the spatio-temporal maps of pulse duration when many fiber modes are excited and when DPC is performed respectively. (e) Envelope of the second order autocorrelation trace of the delivered pulse for the excitation of many fiber modes (black curve, averaged over the camera area) and for the DPC case (blue curve, averaged over the FWHM of the spot size). Dashed red lines are their respective Gaussian fit. The broad pulse (black curve) was scaled to enhance its visibility on the graph. Both pulses possess the same energy. Scale bars are 25 μm. Yellow circles on (a), (b), (c), and (d) indicate the edge of the core of the multimode fiber.

Fig. 5 (a) shows the two-photon signal as a function of optical intensity of the phase conjugated spot of Fig. 4 (b). The non-linear element is a silicon camera which exhibits two-photon absorption at half the wavelength (775 nm) of the 1550 nm excitation. The signal exhibits a quadratic dependence as expected from a two-photon process. Fig. 5 (b) shows the two-photon phase conjugated spot measured with the visible camera. Two-photon absorption suppresses the background intensity, leading to an effective two-photon spot 270 times more intense than its surrounding background. The size of the spot is 5 μm, which is smaller than the 7 μm of the infrared spot shown in Fig. 4 (b) because two-photon absorption occurs in a smaller lateral and axial volume than single photon absorption.

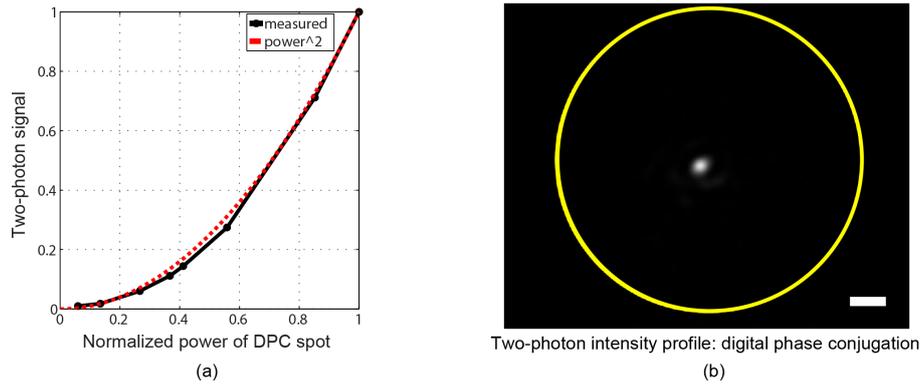

Fig. 5. Two-photon measurement of the phase conjugated spot. (a) Two-photon signal versus normalized power produced by the phase conjugated spot measured on a silicon-based detector (Camera 3 in Fig. 1). Measured (black curve) and theoretical curve (dashed red curve). (b) Two-photon phase conjugated spot. The spot size is 5 μm. The contrast ratio between the maximum intensity to the average background is 270.

## 4. Spot scanning and time multiplexing

*4.1 Scanning of the focused spot*

Following the steps of mode selection explained in section 2, the procedure can be repeated for different locations of the excitation spot at the distal end of the fiber. Then, by using DPC, the field is reconstructed on the proximal side and counter-propagates to generate a focused pulse at each one of those respective locations of the original excitation. The whole process can be considered as a digital scan of the phase conjugated spot. The result is shown in Fig. 6, which presents three spot scanning location captured on Camera 3. The lateral distance between two consecutive spots is 50 μm. The waists of the generated two-photon spots are 5.6 μm, 5.4 μm and 5.9 μm respectively. The intensity of the spot is 48, 135 and 45 times more intense than the background, respectively.

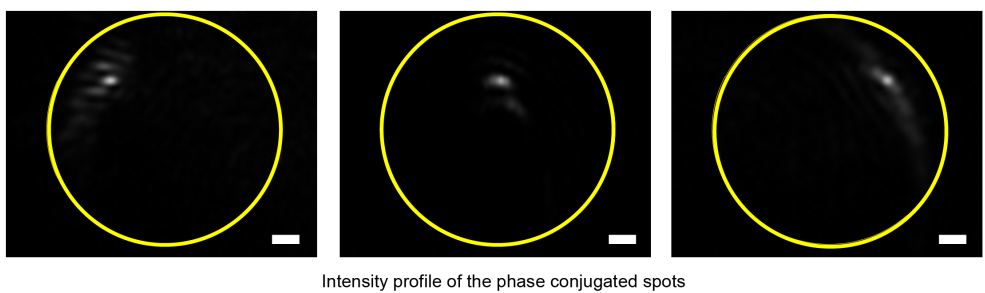

Intensity profile of the phase conjugated spots

Fig. 6. Scanning of the phase conjugated focus. The pulsed intensity focused can be generated at different locations. Scale bars are 25 μm.

*4.2 Time multiplexing*

In addition to the generation of a single phase conjugated spot, we can create a pair (or more) of time delayed spots. This is achieved by reconstructing the superposition of two different groups of modes. The phase pattern on the SLM is simply the sum of the individual phase pattern corresponding to the sampled group of modes. Each group of modes counter-propagates at its own group velocity, arriving to the distal side at two different times, generating two consecutive pulses as shown in Fig. 7. This is possible due to the linearity and time invariance of the propagation of light through a multimode fiber. The pulse duration of each individual pulse is measured by second order autocorrelation. The envelopes of these traces are shown as first pulse and second pulse in Fig. 7. The temporal separation between the pulses is 3.7 ps. Their pulse durations are 500 fs and 800 fs respectively. This experiment demonstrates that it is possible to generate two consecutive high intensity focused pulses at the output of a multimode fiber from a single input pulse.

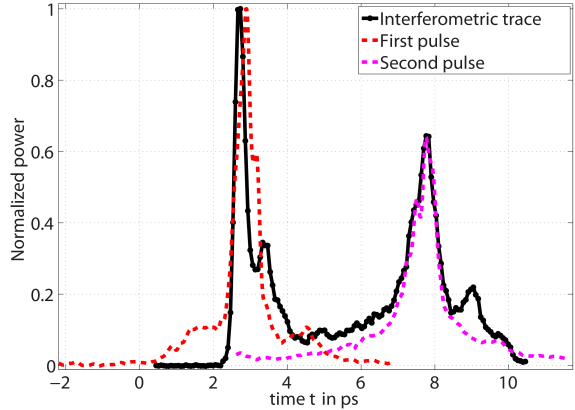

Fig. 7. Generation of two time multiplexed phase conjugated spots. The time envelope is measured by time-gated interferometry (solid curve). The phase conjugated pulses are centered at t=2.9 ps and t=7.7 ps respectively. Dashed curves: second order autocorrelation envelopes of phase conjugated pulses 1 and 2. Their pulse widths are 500 fs and 800 fs respectively.

## 5. Conclusions and outlook

We have presented the first experimental demonstration of simultaneous spatial and temporal focusing of ultrashort pulses through a multimode fiber using time-gated interferometry followed by digital phase conjugation. The focused pulse can be generated at different locations on the distal side of the fiber. Since the power of the phase-conjugated field is spatially spread on the various excited modes, our approach does not introduce nonlinear effects.

We demonstrated how phase conjugation of modes with the same propagation constant reduces modal dispersion in the fiber by a factor of 30. Thus, the pulse preserves its short temporal profile as it propagates. The limited observed temporal spreading (440 fs vs 500 fs) arises due to material dispersion. In our experiments no prechirp was applied to the beam used for the reconstruction of the time-sampled modes. However, if the pulse is prechirped with the right chirp parameter, this temporal spreading could be suppressed.

The ability to focus light through multimode fibers is directly linked to the number of phase conjugated modes. Therefore, the sharpness of the phase conjugated spot could be enhanced by using a multimode fiber with reduced modal dispersion such as a graded-index fiber. We estimate that using a graded index fiber and pre-chirping to correct for material dispersion, the peak intensity of the phase conjugated pulse could be increased two orders of magnitude compared to the step-index multimode fiber case presented here. In fact, by using a graded-index multimode fiber of diameter equal to 200 μm, we would be able to sample and phase conjugate a total of 2200 modes instead of the 42 we have shown, which would significantly increase the contrast of the phase conjugated by a factor of 100.

We also demonstrated how the intrinsic modal dispersion of the multimode fiber can be used to generate two temporally separated pulses from a single pulse. This was accomplished by simultaneous phase conjugation of two set of modes with different group velocities. Each set of modes leads to a single pulse centered at a desired time. In principle this can be extended to a larger number of contiguous pulses.

We have shown that the intensity of the phase conjugated spot is enough to produce two-photon absorption in a silicon-based detector. Therefore, our technique opens the possibility of acquiring two-photon images through commercial multimode optical fibers. Once the system is calibrated, a two-photon fluorescence image of a sample placed in the distal side of the fiber can be obtained by scanning the DPC spot over the sample and collecting the fluorescent signal through the same fiber.

**Appendix A: Intensity of the phase-conjugated spot**

For efficient non-linear imaging such as two-photon imaging through a multimode optical fiber, a focused spot with the smallest size and the highest peak intensity is required. The spot size and its temporal duration depend strongly on the set of modes excited to generate it. Indeed, high order modes in an optical fiber are associated with a higher numerical aperture than lower order modes. Thus, the spot size of the phase conjugated spot is expected to decrease with higher mode selection, as experimentally demonstrated in Fig. 3 (a-d) and (f). We define the peak intensity of the phase conjugated spot as:

$$I_{peak} = \frac{E_{spot}}{\sigma s \, \sigma t}$$

where $E_{spot}$, $\sigma s$ and $\sigma t$ are the energy, the area, and the temporal duration, respectively, of the phase conjugated spot.

We seek a phase conjugated spot that produces the largest peak intensity. Hence, by measuring the energy, spot size and temporal duration for different phase conjugated set of modes, we obtain the peak intensity dependence on time $\tau_1$ shown in Fig. 8.

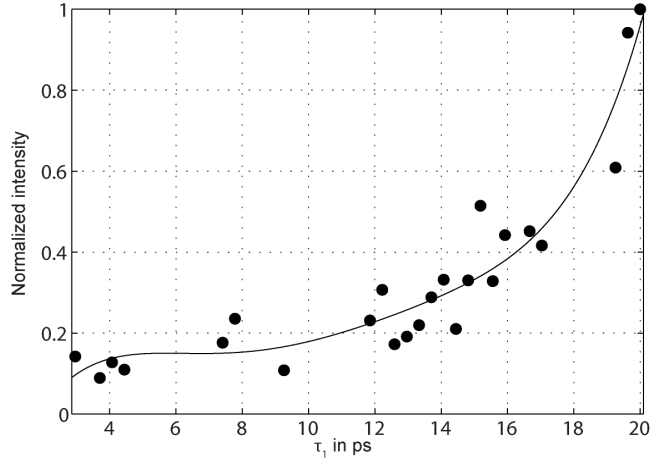

Fig. 8. Normalized peak intensity of the phase conjugated spot versus time $\tau_1$. Points represent experimental data and the solid lines represent their polynomial fit. The optimal time interval to phase conjugate can be therefore chosen based on the maximized intensity given by the dataset.

We observe that higher order modes provide the largest peak intensity of the phase conjugated spot at the distal end of the fiber. In fact, the highest peak intensity measured in our experiment was 125 MW/cm$^2$ realized with a spot waist of 5.1 μm corresponding to the excitation of high order modes sampled at a delay $\tau_1 = 20$ ps.